\documentclass[10pt,twocolumn,prl,nofootinbib,preprintnumbers,amssymb,amsfonts,amsmath,superscriptaddress,showpacs]{revtex4-1}

\usepackage{graphicx}
\usepackage{bm}
\usepackage{color}
\usepackage{subfigure}
\usepackage{epstopdf}

\def\simleq{\; \raise0.3ex\hbox{$<$\kern-0.75em \raise-1.1ex\hbox{$\sim$}}\; }
\def\simgeq{\; \raise0.3ex\hbox{$>$\kern-0.75em \raise-1.1ex\hbox{$\sim$}}\; }

\newcommand{\GeV}{{\rm GeV}}
\newcommand{\TeV}{{\rm TeV}}

\newcommand{\kpc}{{\rm kpc}}

\begin{document}

\title{A common solution to the cosmic ray anisotropy and gradient problems}

\author{Carmelo Evoli}
\email{carmelo.evoli@desy.de}
\affiliation{{II.} Institut f\"ur Theoretische Physik, Universit\"{a}t Hamburg, LuruperChaussee 149, D-22761 Hamburg, Germany}

\author{Daniele Gaggero}
\email{daniele.gaggero@pi.infn.it}
\affiliation{INFN Pisa and Pisa University, Largo B. Pontecorvo 3, I-56127 Pisa, Italy}
\affiliation{LAPTh, Univ. de Savoie, CNRS, B.P.110, Annecy-le-Vieux F-74941, France}

\author{Dario Grasso}
\email{dario.grasso@pi.infn.it}
\affiliation{{II.} Institut f\"ur Theoretische Physik, Universit\"{a}t Hamburg, LuruperChaussee 149, D-22761 Hamburg, Germany}
\affiliation{Dipartimento di Fisica, Universit\`a di Siena, Via Roma 56, I-56100 Siena, Italy}

\author{Luca Maccione}
\email{luca.maccione@lmu.de}
\affiliation{Ludwig-Maximilians-Universit\"{a}t, Theresienstra{\ss}e 37, D-80333 M\"{u}nchen, Germany}
\affiliation{Max-Planck-Institut f\"{u}r Physik (Werner Heisenberg Institut), F\"{o}hringer Ring 6, D-80805 M\"{u}nchen, Germany}

\begin{abstract}
Multichannel Cosmic Ray (CR) spectra and the large scale CR anisotropy can hardly be made compatible in the framework of conventional isotropic and homogeneous propagation models. These models also have problems explaining the longitude distribution and the radial emissivity gradient of the $\gamma$-ray galactic interstellar emission. We argue here that accounting for a well physically motivated correlation between the CR escape time and the spatially dependent magnetic turbulence power can naturally solve both problems.  Indeed, by exploiting this correlation we find propagation models that fit a wide set of CR 
spectra, and consistently reproduce the CR anisotropy in the energy range $10^2 - 10^4~\GeV$ and the $\gamma$-ray longitude distribution recently measured by Fermi-LAT. 
 \end{abstract}

\maketitle 

\noindent{\underline{\em Introduction}:} 
The propagation of Cosmic Rays (CR) throughout the Galaxy is far from being fully understood. 
One of the most important, and still uncertain, quantities which determine the CR spectra and spatial distribution in the Galaxy is the power-law spectral index ($\delta$) of their rate of escape as a function of energy.  This quantity is typically probed by secondary/primary CR ratios, the Boron to Carbon ratio (B/C) most commonly, which is measured with high accuracy up $\sim 1~\TeV$.  
Several degeneracies with other physically relevant processes, however, prevent from casting stringent constraints on $\delta$. Indeed, depending on the adopted values of other (poorly known) parameters as the Alfv\'en and convective velocities, we can have $\delta \sim 0.5 - 0.6$  for plain diffusion (PD) and low reacceleration models (see e.g.~\cite{Jones:2000qd,Ptuskin:2005ax,DiBernardo:2009ku}), $\delta \sim 0.8-0.9$ for convective models \cite{Maurin:2010zp} or  $\delta \sim 0.3$ for non-convective - strong reacceleration models \cite{Trotta:2010mx}.
In principle, measurements taken in other CR channels could help reducing some of those degeneracies. However, multi-channel analysis opens new problems. For example, strong reacceleration -  low $\delta$ models are disfavored by antiproton data (e.g.~\cite{DiBernardo:2009ku,Strong:2007nh} and Ref.s therein). The observed low energy CR electron and positron spectra \cite{DiBernardo:2010is} and the spectrum of the synchrotron emission of the Galaxy \cite{Strong:2011wd} are also hardly compatible with those models. Furthermore models with small $\delta$ are in tension with the modeling of CR acceleration in sources as they require too steep source spectra in order to reproduce the observed propagated CR spectra \cite{Caprioli:2011ze}. On the other hand, models with $\delta \gtrsim 0.5$, which are preferred on the basis of CR spectral data, face major problems  with the observed large scale CR anisotropy. 

Diffusion of charged CRs in the turbulent Galactic Magnetic Fields (GMFs) erases almost completely the information about source positions, leaving only a small residual anisotropy. 
Current data (see {\it e.g.}~\cite{Guillian:2005wp} and ref.s therein) show a dipole-like large scale anisotropy (LSA) of the order of $10^{-3}$ very weakly energy dependent above $\sim10~\TeV$, plus several smaller scale anisotropies in various directions. Below few TeV the observed LSA depends more strongly on energy and drops down to $\sim10^{-4}$ at 100 GeV. 
Apart the Compton-Getting effect (yielding an energy independent anisotropy at the $\sim 10^{-4}$ level), the origin of the observed anisotropy can be twofold: the CR drift due to the inhomogeneous source distribution and the global leakage from the Galaxy, and the stochastic effect of local  sources.
While source stochasticity can likely explain the observed behavior of the anisotropy above 10 TeV \cite{Ptuskin:2006}, it cannot reconcile the anisotropy data with the LSA predicted by diffusive models with $\delta \gtrsim 0.5$ \cite{Blasi:2011fm}. 
 We refer to such discrepancy as the {\em CR anisotropy problem}. 

Non-local observables, like the galactic $\gamma$-ray interstellar emission, are a precious probe of the large scale CR spatial distribution, hence they may shed light on the origin of the anisotropy problem. 
The interpretation of the angular distribution of that emission in terms of standard diffusion models also faces a long withstanding problem.  It has been known since the EGRET era \cite{Hunger:1997we} that the CR galactocentric radial distribution, derived assuming a source density deduced from pulsar or supernova remnant (SNR) catalogues, is much steeper than the one inferred from the $\gamma$-ray interstellar emission along the Galactic plane.
This is known as the {\em CR gradient problem}. 

Thanks to its high angular resolution, Fermi-LAT, upon recently confirming the existence of the gradient problem \cite{collaboration:2009ag,Collaboration:2010cm}, allowed also to separate the emission coming from CR interactions with the molecular gas (whose modeling is strongly affected by the uncertainty on the conversion factor between the tracer CO intensity and the ${\rm H_2}$ column density, ${\rm X_{CO}}$) from the emission due to CR interactions with the atomic gas (whose density is better known from its 21 cm radio emission). An analysis based on $\gamma$-ray maps of the third Galactic quadrant \cite{Collaboration:2010cm} (see also \cite{Collaboration:2012uz}) showed that the emissivity from {\it neutral} gas shows a radial decrease smaller than the one predicted by conventional models. This confirms the gradient problem independently of the uncertainties on the ${\rm X_{CO}}$ parameter, thereby strongly disfavoring explanations invoking a sharp rise of the X$_{\rm CO}$ in the outer Galaxy 
\cite{Strong:2004td}. Some alternative explanations explored in \cite{Collaboration:2010cm} do not appear satisfactory: a thick halo is disfavored both by $^{10}$Be/$^9$Be \cite{Strong:2007nh} and radio data (see e.g.~\cite{Bringmann:2011py} and ref.s therein); a smooth source distribution is in contrast with SNR catalogues. 
Furthermore, because the LSA is proportional to the diffusion coefficient (DC) $D$ (see below) and $D/H$ (where $H$ is the diffusive halo half-height) is fixed against the B/C ratio, a thick halo would worsen the anisotropy problem. 
  
In this letter we argue that the CR gradient and anisotropy problems share the same origin, residing in a too simplified treatment of CR diffusion commonly assumed to be isotropic and spatially homogeneous.  
We drop both these assumptions (supported neither by theoretical arguments nor by numerical simulations)
and show that, by making CR escape faster in the most active regions, hence the CR radial distribution smoother, we solve naturally both the CR gradient and anisotropy problems. 
We exhibit physically motivated propagation models that, while fitting local data of CR spectra, yield also the radial CR distribution required to reproduce the diffuse $\gamma$-ray emission measured by Fermi-LAT, similarly to what we already found to interpret EGRET data \cite{Evoli:2008dv} (see also \cite{Breitschwerdt:2002vs,Gebauer:2009hk} for other possible approaches).
Moreover we  show for the first time that these models predict a significantly smaller CR anisotropy than the corresponding models assuming isotropic and uniform diffusion.  As a consequence, the agreement with the measured anisotropy is very much improved, especially in the energy range $0.1\div10$ TeV, hinting at a new self-consistent and comprehensive description of present CR data. 

\noindent{\underline{\em Anisotropic inhomogeneous diffusion}:} 
Charged CRs diffuse in the turbulent component of the GMF. The regular GMF (directed almost azimuthally along the spiral arms) breaks the isotropy hence the diffusion tensor has to be expressed in terms of two coefficients $D_\parallel$ and $D_\perp$ describing diffusion in the parallel and perpendicular direction to the regular GMF.  For weak turbulence $\eta \equiv \delta B/B \ll 1$ ($\delta B$ represents the rms fluctuation of the magnetic field) quasi linear theory (QLT) predicts $D_\perp \approx D_\parallel \eta^2$, hence perpendicular diffusion is irrelevant in that case; in the opposite case isotropy should be restored. 

In the Galaxy, however, where the regular and the turbulent components have similar strengths $(\eta \simeq 1)$ the situation is more complex and can be reliably studied only by means of numerical simulations. 
Here we refer to the results reported in \cite{DeMarco:2007eh}, where propagation of CR protons in simulated realizations of random MFs  with $\eta = 0.5 \div 2$ and in the presence of a large scale azimuthal field was discussed. 
For the random component, a Kolmogorov spectrum was assumed in agreement with observations \cite{Elmegreen:2004wj}. 
Simulations were performed only for proton energies above $10^{15}$ eV ($10^{14}$ eV for $\eta = 1$) due to computer time limitations. Since the simulated values of $D_\parallel$ and $D_\perp$ decrease slowly and steadily with decreasing energy, and there are no reasons why such behavior should change at lower energies, we assume that their ratio can reliably be extrapolated down to the energies considered in this work.  As we will show, this assumption is indeed consistent with several experimental facts. 
Remarkably, only the simulated value of the $D_\perp/D_\parallel$ ratio is relevant here since the absolute values of the DC components will be fixed against CR data. 

Similarly to QLT predictions, the authors of \cite{DeMarco:2007eh} found $D_\parallel$ and $D_\perp$ to behave oppositely with respect to the turbulent power, the former (latter) decreasing (increasing) with $\eta$.  
This can be understood as GMF lines random walk being enhanced with increasing turbulence strength. 
Moreover, they found that while $D_\parallel \propto E^{1/3}$ (in agreement with QLT)  the slope of $D_\perp$ is significantly steeper: $D_\perp \propto E^{0.5 \div 0.6}$. 
Perpendicular diffusion should thus be the dominant CR escape channel in the inner Galaxy and in the next outer regions \cite{DeMarco:2007eh}. Indeed, 
for $H < R_{\rm Gal} - R_\odot$, with $R_{\rm Gal} \simeq 20~\kpc$ and $R_\odot \simeq 8.5~\kpc$,  the parallel/perpendicular escape time ratio is
\begin{equation}
\frac {T_\parallel}{T_\perp} \simeq \left( \frac{R_{\rm arm}}{H}\right)^2~\frac  {D_\perp} {D_\parallel}\simeq 4 \times 10^2  \left( \frac{H}{4~ \kpc}\right)^{-2}~  \frac{D_\perp} {D_\parallel} 
\end{equation}
where $R_{\rm arm} \simeq \pi R_\odot$  
is roughly the length CR have to travel to escape the Galaxy along the spiral arms, parallel to the regular GMF. 
For $\eta \sim 1 - 2$, 
$D_\perp/D_\parallel \gtrsim 10^{-2}$ was found at $10^5~\GeV$ slowly decreasing with decreasing energy \cite{DeMarco:2007eh}.  Extrapolating this result down to $\sim 10~\GeV$,  we still find  $T_\perp < T_\parallel$ unless $H$ is considerably larger than the preferred value $H \simeq 4~\kpc$ determined on the basis of the observed $^{10}{\rm Be}/^{9}{\rm Be}$ ratio and radio data.

The above considerations have two main consequences:
1) Under the observationally preferred conditions of Kolmogorov turbulence,  the CR escape time is expected to depend on energy as $E^{0.5 - 0.6}$. Noticeably, this is the same dependence which is favored by a combined analysis of CR nuclei and antiproton spectra \cite{DiBernardo:2009ku}.  
2) In the inner Galaxy at least,  $T_{\rm esc}  \sim  T_\perp = {H}^2/6 D_\perp$ should be anti-correlated to the turbulent power, hence to the density of CR sources  (see also \cite{Evoli:2008dv}), which we assume inject turbulence in the ISM.

\noindent{\underline{\em Solution of the CR gradient problem}:} 
We study the effects of our assumption of a spatial correlation between the DC and the source density by solving the diffusion equation with the {\tt DRAGON} numerical diffusion code \cite{dragonweb}, which, differently from other numerical and semi-analytical programs,  is designed to account for a spatially dependent DC.  
The code is 2-dimensional ($R, z$) and assumes a purely azimuthal (no arms) structure of the regular GMF. Therefore we can only model perpendicular diffusion and the DC is treated as a (position dependent) scalar. Nevertheless, as only the escape time is relevant to determine the CR density, we can account for parallel diffusion along the spiral arms 
by using an effective DC:  $D_{\rm eff}(R) = {\rm max}\left[ D_\perp(R), (H/R_{\rm arm})^2~D_\parallel(R) \right ]$. 
We assume therefore the phenomenological dependence $D_\perp(R) \propto Q(R)^{\tau}$, where $\tau \simgeq 0$ is a free parameter to be fixed against data (simulations do not allow to determine $\tau$ with sufficient accuracy). According to QLT and numerical simulations we assume $D_\parallel$ to have an opposite dependence on the turbulence strength, hence $D_\parallel(R) \propto Q(R)^{- \tau}$. 
We remark that parallel diffusion has almost no effect on the $\gamma$-ray angular distribution and the local CR anisotropy, as it becomes relevant only in the most external regions of the Galaxy, where the source density (hence turbulence injection) is very small. Its presence, however, naturally prevents the escape time from taking unphysical large values at large $R$.
 For the source radial distribution we adopt $Q(R) \propto (R/R_\odot)^{1.9}\exp(- 5 (\frac{R - R_\odot}{R_\odot}))$, based on pulsar catalogues \cite{Lorimer:2006qs}.  Using other, observationally determined, distributions would not change our main results. 
Similarly to \cite{DiBernardo:2009ku,Evoli:2008dv} we assume a vertical profile $D_{\rm eff}(R,z) = D_{\rm eff}(R)\exp{(z/H)}$.  We also assume $D \propto (v/c)^{-0.4}$ 
($v$ is the particle velocity)  to reproduce the low-energy B/C data as shown in those papers.  This does not affect the results discussed here.
We fix $H = 4~\kpc$ and for each value of $\tau$ we set the $D$ normalization to match the observed B/C and other light nuclei ratios. 
We fix the $D$ rigidity dependence  $\delta = 0.6$ in the rest of our Letter.
To better highlight the effects of inhomogeneous diffusion we consider here only PD propagation setups. Adding moderate reacceleration and radially uniform convection does not change significantly any of our results. 

\begin{figure}[tbp]
\centering
\includegraphics[width=0.45\textwidth]{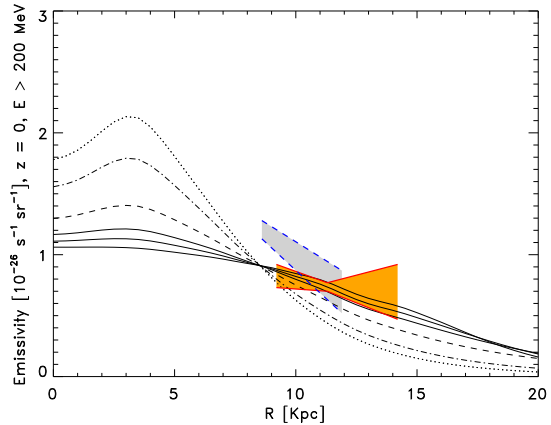}
\caption{Integrated $\gamma$-ray emissivity (number of photons emitted per gas atom per unit time) constrained by Fermi-LAT (orange region \cite{Collaboration:2010cm}, grey region \cite{collaboration:2009ag}) compared with our predictions for  $\tau=0,0.2,0.5,0.7,0.8,0.9$ (from top to bottom).
}
\label{fig:PD_exprad_00_02_05_07_08_10_Rmax30}
\end{figure}

\begin{figure}[tbp]
\centering
\includegraphics[width=0.45\textwidth]{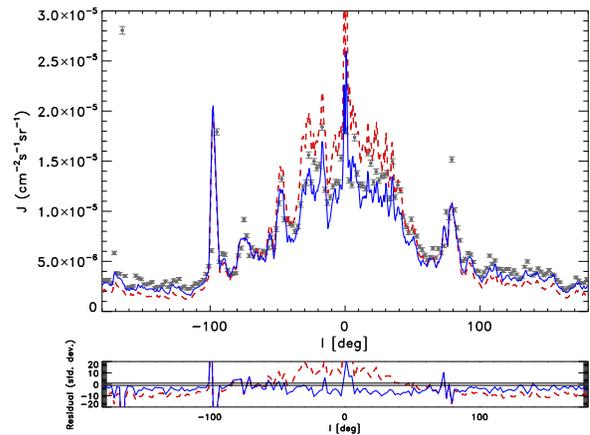}
\caption{Predicted longitudinal profile of the $\gamma$-ray diffuse flux along the Galactic plane compared to Fermi-LAT data \cite{Strong:2011pa}, and residuals.   Data are integrated over the latitude interval $|b| < 5^{\circ}$ and in energy between 1104 and 1442 MeV. Solid (blue) line $\tau = 0.85$, dashed (red)  line $\tau=0$. }
\label{fig:PD_D_exp_exprad}
\end{figure}

We find a good fit of the B/C for all values of $\tau\in[0,1]$. The best fit $D$ normalization only mildly depends on $\tau$. 
Also the computed antiproton and mid-latitude $\gamma$-ray spectra match observations within errors. 
We then calculate the $\gamma$-ray emissivity from the CR spatial distributions in our models. 
As clear from Fig.~\ref{fig:PD_exprad_00_02_05_07_08_10_Rmax30}, the model $\tau = 0$ (uniform diffusion) does not reproduce the observed emissivity profile. 
We obtain the simulated $\gamma$-ray angular distribution by performing a line-of-sight integration of the product of the emissivity times the gas density. For consistency we use the same gas distribution \cite{galpropweb} and the same catalogue sources \cite{Strong:2011pa} adopted by the Fermi-LAT collaboration.   
We show in Fig.~\ref{fig:PD_D_exp_exprad} the longitude profiles of Galactic $\gamma$-ray emission and the residuals of the models against data for $\tau=0$ and $\tau=0.85$.
The model $\tau = 0$ is clearly too steep compared to data:  it overshoots the data in the Galactic center region while it undershoots observations by several $\sigma$  in the anti-center region.  
Increasing $\tau$ yields a much smoother behavior of the emissivity as function of $R$ (see \cite{Collaboration:2010cm} for the possible reasons why the emissivity in the II and III quadrants do not agree entirely). A good match of Fermi-LAT data is achieved for $\tau \simeq {\rm [0.7 \div 0.9]}$, with $\tau = 0.85$ providing an optimal fit and improving the residual distribution.	
\noindent{\underline{\em Effect on the CR anisotropy:}} 
The CR LSA component in the radial direction is related to the CR gradient by
\begin{equation}
{\rm anisotropy} = \frac{3D_{\perp}}{c}  \left\vert{  \frac{\nabla_{r} n_{\rm CR}}{n_{\rm CR}}}\right\vert,
\label{eq:anisotropy}
\end{equation}
which we use to compute the contribution of CR diffusion to the LSA starting from the CR distribution computed in the same PD models as in the previous section. 
Remarkably, with increasing $\tau$, hence with a smoother CR distribution, the predicted LSA also decreases.  Changing from $\tau = 0$ to $\tau = 1$ reduces the anisotropy by almost a factor of 10. 
Intriguingly, we can reproduce the CR anisotropy data \cite{Guillian:2005wp} up to few $\TeV$ with $\tau = 0.85$. 
\begin{figure}[tbp]
\centering
\includegraphics[width=0.45\textwidth]{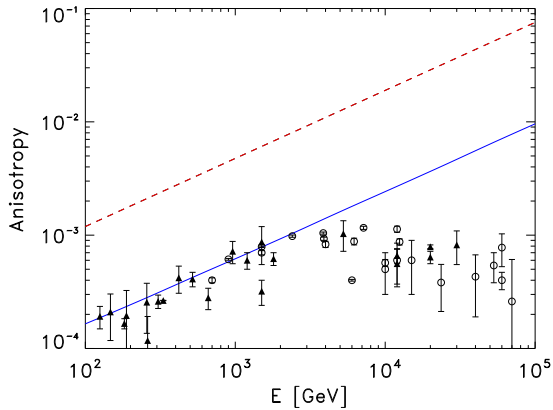}
\caption{The CR anisotropy measured by several experiments is compared with our predictions for $\tau=0$ (dashed red line) and $\tau = 0.85$ (solid blue line).
Triangle/circle data were taken from muon/EAS detectors as reported in \cite{Guillian:2005wp} and \cite{DiScascio:2012br}.}
\label{fig:anisotropy_pd}
\end{figure}
The discrepancy between our model and the observed anisotropy above that energy is probably due to source stochasticity which we did not account for in this work.  
Indeed, while below 10 TeV the observed anisotropy phase (see \cite{Guillian:2005wp} and ref.s therein) keeps almost constant to a value compatible with expectations from the global CR leakage, above that energy it significantly fluctuates, as expected if the contribution of stochastic sources becomes dominant.  
 
\noindent{\underline{\em Conclusions}:}
In this letter we presented a consistent solution to the CR gradient and anisotropy problems. Our approach is based on the physically motivated hypothesis that the CR diffusion coefficient is spatially correlated to the source density: regions in which star, hence SNR, formation is stronger are expected to show a stronger turbulence level and therefore a larger value of the perpendicular DC (oppositely to what happens for $D_\parallel$). The escape of CRs from most active regions is therefore faster, hence smoothing out their density through the Galaxy. Correspondingly, the predicted CR gradient and anisotropy are reduced. 
We implemented a phenomenological realization of this scenario and checked that -- while CR data are still correctly reproduced -- our approach also gives a remarkably good description of the spectrum and longitude distribution of the diffuse $\gamma$-ray emission measured by the Fermi-LAT collaboration. Our analysis provides for the first time a unified propagation model which reproduces local nuclear spectra and also explains non-local observables, and in particular reconciles the preferred low-reacceleration models with $\delta\simeq0.5$ hinted at by the combined spectra of nuclei (B/C), antiprotons, electrons and radio data (and phenomenologically preferred by acceleration theory) with anisotropy and gradient observations. We take these results as an encouragement to pursue a self-consistent theory/computation of non-linear CR - MHD turbulence interaction in the Galaxy. We notice that an alternative solution of the CR gradient problems in terms of a spatially varying convective velocity was proposed in \cite{Breitschwerdt:2002vs,Gebauer:2009hk}. A possible consistent solution of CR isotropy problem also deserves to be investigated.    

\noindent{\underline{\em Acknowledgments}:}
We warmly thank P.~Blasi, A.~Strong and L.~Tibaldo for reading the draft of this paper and providing useful insights. 
We also thank G. Di Sciascio for kindly providing us with CR anisotropy data.  
CE acknowledges support from the Helmholtz Alliance for Astroparticle Phyicsâ funded by the Initiative and Networking Fund of the Helmholtz Association. The work of D.~Grasso is supported by the DFG through the collaborative research centre SFB 676. 
LM acknowledges support from the AvH foundation.

\end{document}